\documentclass[11pt]{article}
\usepackage[centertags]{amsmath}
\usepackage{amsfonts}
\usepackage{latexsym,amssymb}
\usepackage{graphics}
\newcommand{\pr}{\mbox{\sf P}}
\newcommand{\ex}{{\bf\sf E}}               
\newcommand{\var}{\mbox{\sf Var}}

\newcommand{\by}{{\bf y}}               

\newcommand{\al}{\alpha}                
\newcommand{\g}{\lambda}                
\newcommand{\G}{\Lambda}                
\newcommand{\lam}{\lambda}               
\newcommand{\Lam}{\Lambda}               

\newcommand{\startb}{\parindent0pt\bf}  

\newtheorem{thm}{Theorem}
\newtheorem{lem}[thm]{Lemma}
\newtheorem{pro}[thm]{Proposition}

\newtheorem{exm}{Example}
\newtheorem{rem}{Remark}

\def\th{\theta}
\def\Th{\Theta}

\def\nd{\quad{\rm and}\quad}

\begin{document}

\baselineskip 18pt

\title{\bf The Stochastic Knapsack Revisited:  \\
Switch-Over Policies and Dynamic Pricing}

\author{Grace Y.\ Lin, Yingdong Lu \\
IBM T.J.\ Watson Research Center\\ Yorktown Heights, NY 10598 \\
E-mail: \{gracelin, yingdong\}@us.ibm.com \\
\\
David D. Yao\thanks{
Research undertaken while an academic visitor
at IBM T.J.\ Watson Research Center; also
supported in part by NSF grant DMI-0085124 and
Hong Kong RGC Grant CUHK4173/03E.}\\
IEOR Dept., 
Research \\
Columbia University\\
New York, NY 10027\\
E-mail: yao@ieor.columbia.edu}

\date{(revision: April 2006)}

\maketitle

\begin{abstract}
The stochastic knapsack has been used as a model in
wide ranging applications from
dynamic resource allocation to
admission control in telecommunication.
In recent years, a variation of the model has become a
basic tool in studying problems that arise in revenue management
and dynamic/flexible pricing; and it is in this context that
our study is undertaken.
Based on a dynamic programming formulation and
associated properties of the value function, we
study in this paper a class of control that we call switch-over
policies
--- start from accepting only orders of the highest price,
and switch to including lower prices as time goes by, with
the switch-over times optimally decided via convex programming.
We establish the asymptotic optimality of the switch-over policy,
and
develop pricing models based on this policy
to optimize the price reductions over the decision horizon.
\end{abstract}

\section{Introduction}
\label{sec:introduction}

The {\it stochastic knapsack} refers in general to a dynamic
resource allocation problem in which a fixed amount of resource is
allocated sequentially to random demands of multiple classes. The
problem appears to have many different roots. In Derman {\it et
al} (1972), a Markov decision problem is formulated to determine
how to assign fixed amount of different resources to sequential
arrivals with random unit returns. In Prastacos (1983), a problem
for sequential investment is analyzed which allows only one
acceptance decision to be made over the horizon. In the late
1980's, the stochastic knapsack was used as a model to study
admission control in telecommunication networks, so-called loss
networks in particular; refer to Ross and Tsang \cite{keith}, and
Ross and Yao \cite{rossyao}.

In recent years, one version of the stochastic knapsack has become
a basic model in studying problems in the general area of revenue
management. Here the capacity of the
knapsack corresponds to a given amount of resource that can be
used to fulfill customer demand, over a given time frame that is
typically quite short. Some examples include: rooms in a hotel
targeted for weekend tourists, seats on an airplane that must be
sold before departure, fashion items at a department store
designed for a particular season. 
Refer to Kleywegt and Papastavrou
\cite{kleywegt2,kleywegt3}, Papastavrou {\it et al.}
\cite{kleywegt1}, and Van Slyke and Young \cite{vanslyke}.

Our study here also falls into this category, although it is
motivated by a new application. A major producer of personal
computers (PC's) from time to time has to liquidate a substantial
inventory of desktop or laptop computers returned from corporate
leases. This involves a number of sales channels --- including
catalog direct sales, dealers/brokers, and on-line auction ---
which differ in both price and batch size. What is needed by
practitioners involved  in this process is a tool to support
pricing decisions. Specifically, not only what type of ``bids''
(customer demand or offer to buy the PC's) to accept and when (in
terms of time and available inventory), but also how to price the
sales over time.
  Refer to more details in \cite{flexiblepricing,caojanglu}.

As evident from previous studies, with prices given, the problem
of when to accept or reject which demand can be formulated and
solved using dynamic programming (DP). This, however, does not
in general lead to optimal policies that have simple and easy-to-implement
structures; neither does it result in a tractable value function,
upon which pricing optimization can be carried out.
Our approach makes use of certain properties of the DP value
function,
such as concavity and submodularity (which have been
familiar properties in other application contexts such as
queueing control, e.g., Lippman \cite{lippman}),
which lead to a lower- and upper-orthant structure of the optimal
policy (see Proposition \ref{pro:structure}). This structure, in turn,
motivates us to focus on a class of ``switch-over'' policies.

The switch-over policy is executed in the following manner: it
starts from accepting only orders of the highest price, and
gradually switches to including lower prices as time goes by.
There are several advantages in focusing on this class of
policies: a) they are consistent with the lower- and upper-orthant
structure associated with the optimal policy; b) they are easy to
identify: the optimal switch-over times are readily derived
through convex programming; and most importantly, c) they are
practical for implementation: indeed they follow closely the
prevailing practice in certain industry sectors. Furthermore, we
can prove that the switch-over policy, while sub-optimal in
general, is asymptotically optimal in the sense that the relative
error between the switch-over and the optimal policies goes to
zero as the available inventory increases to infinity (along with
the planning horizon).

One aspect of our work that is a departure from previous
studies is that we are concerned with setting the optimal prices
(in addition to deciding order acceptance/rejection). Based on the
switch-over policy, we formulate optimization problems, so that
the reduced (``sales'')  prices over the decision horizon can be
optimally determined, taking into account that the rate of demand
is a (decreasing) function of the price. A closely related set of
papers in revenue management, although not always making an
explicit connection to the stochastic knapsack, studies problems
that are similar to ours in both physical and mathematical
aspects. These include Bitran and Mondschein \cite{bitran},
Brummelle and Walczak \cite{brum}, Feng
and Gallego \cite{fenggallego1995,fenggallego2000}, Feng and Xiao
\cite{fengxiaoor2000}, Lee and Hersh \cite{lee},
and Zhao and Zheng \cite{zhaozheng}. 

Briefly, the rest of the paper is organized as follows. In
\S\ref{sec:problem},
we start with a dynamic programming problem
formulation, and bring out the structure of the optimal policy.
In the next two sections,
we focus on the switch-over policy,
starting with the case of constant batch sizes in \S\ref{sec:switch}.
The general case of random batch
sizes is studied in \S\ref{sec:batch}, and
the asymptotic optimality of the switch-over policy is established in
\S\ref{sec:asymptotic}. We then develop the pricing models in
\S\ref{sec:pricing}, and conclude with possible extensions in
\S\ref{sec:conclude}.

\section{The Dynamic Programming Formulation}
\label{sec:problem}

Here is a formal description of our model.
There are $W$ units
of inventory available to supply the orders at times $n=1,\cdots,
T$, where $T$ is a given integer, representing the planning
horizon. The order (demand) that arrives in each period $n$
takes the form of a bivariate random vector: $(P_n, Q_n)$, where
the two components represent the unit offer price and the required
quantity. Suppose $(P_n,Q_n)$ are i.i.d.\ across
$n$, following a joint distribution:
\begin{eqnarray}
\label{jointdistn}
\pr [P_n=p_i, Q_n=j]:=\th_{ij}, \qquad
i=1,...,m; \quad j=1,...,W .
\end{eqnarray}
Here, we assume, for all $i$ and $j$,
$p_i >0$ and
$\sum_i\sum_j \th_{ij} \le 1$,
with
$$\th_0:=1-\sum_i\sum_j \th_{ij} \ge 0$$
representing the
probability that there is no order arrival in a period.

In each period, our decision is, after observing the realized
$(P_n,Q_n)$, whether or not to supply the order. If we do, a
revenue
of $P_nQ_n$ is collected; otherwise, we earn nothing. Here, we
assume that each order is either supplied in
full, or not at all; i.e., no partial supply is allowed.
In particular, if
the inventory available upon an order arrival is less than the
order size, then no supply takes place. The objective is
to maximize the expected revenue collected over the planning
horizon of $T$ periods and the total available inventory of $W$
units.

Let $V(n,d)$ denote the expected revenue we can collect, under
optimal actions, starting from period $n$ ($\le T$), with $d$
($\le W$) units of inventory left. Then, we have the
following dynamic programming (DP) recursion:
\begin{eqnarray}
\label{dprecursion}
V(n,d)&=&V(n+1,d)[\th_0+\Th (d)]\nonumber\\
&+&\sum_{i}\sum_{j\le d}\th_{ij}\cdot
\max\{p_ij+V(n+1,d-j), V(n+1,d)\},
\end{eqnarray}
where
\begin{eqnarray}
\label{Th}
\Th(d):=\sum_{i}\sum_{j> d}\th_{ij}.
\end{eqnarray}
Clearly, the first term on the right hand side of
(\ref{dprecursion}) corresponds to the case of either no arrival
or the order size exceeds the available inventory; whereas each
term under the double summation compares the two actions: accept
(i.e.,
supply) the order, or reject it. If we supply the order, then we
earn the revenue $p_jj$, and proceed to the next period with
$j$ units less in the available inventory.
In the last period, we have
\begin{eqnarray}
\label{boundary}
V(T,d)=\sum_{i}\sum_{j\le d}\th_{ij} p_ij,
\end{eqnarray}
since clearly the best action is to supply any possible order
using all the remaining inventory.

The above DP is quite easy to solve -- the overall
computational effort is, after all, only $O(TW)$.
Short of any structural properties, however, the solution
does not readily translate into
a policy that is easy to implement.
We need to
pre-compute and store the $V(n,d)$ values for all $n=1,...,T$ and
all $d=1,...,W$. Then, after observing the realized demand
$(P_n,Q_n)$, we will supply it, if $Q_n\le d$ and
\begin{eqnarray}
\label{opt_threshold}
 P_nQ_n +V(n+1,d-Q_n)\ge V(n+1,d);
\end{eqnarray}
and reject it otherwise.
Furthermore, the lack of a tractable form of the
value function with respect to the prices is a severe
handicap when it comes to solving the optimal pricing problem.

In the case that all orders are of unit size, 
the above problem can be reduced to a special case of 
a certain queueing control model
in Lippman \cite{lippman}, and
the value function, $V(n,d)$, can be shown to be
concave in $d$ and submodular in $(n,d)$.
These properties have the following implications:

\begin{itemize}
\item
For each price type $i$,
if the order is rejected in some state
$(n^*,d^*)$, then it is rejected in all ``lower'' states $(n,d)\le
(n^*,d^*)$;
if the order is accepted in some state
$(n^*,d^*)$, then it is accepted in all ``upper'' states $(n,d)\ge
(n^*,d^*)$.
\end{itemize}

To understand the above, we know from the DP recursion, 
if a type $i$ order is rejected in state $(n^*,d^*)$, then
$$p_i\le V(n^*,d^*)-V(n^*,d^*-1).$$
For any state $(n,d)\le (n^*,d^*)$, we have
$$V(n^*,d^*)-V(n^*,d^*-1)\le
V(n^*,d)-V(n^*,d-1)\le V(n,d)-V(n,d-1),$$
due to the concavity (the
first inequality) and submodularity (the second inequality) of $V$.
Hence, 
$$p_i\le V(n,d)-V(n,d-1),$$
i.e., the same order should also be rejected in state $(n,d)$ as well, which
is the lower-orthant property.
On the other hand, if the order is accepted in state $(n^*,d^*)$,
then it must be accepted in any state $(n,d)\ge (n^*,d^*)$; for
if it is rejected in state $(n,d)$, then
following the lower-orthant property,
it must also be rejected
in state $(n^*,d^*)$.

The above properties, stated in a slightly different but equivalent form
below, are the basis
for the switch-over policies that we shall focus on in the rest of
this paper.
 
\begin{pro}
\label{pro:structure} 
{\rm Suppose all orders are of the
same, constant size. (Hence, without loss of generality, assume
this constant size is unity,
 i.e., $Q_n\equiv 1$.) Then,
the  optimal policy has the following structure:
 for each inventory level $d$, there exist time epochs $0=t_0(d)\le t_1(d) \le
\cdots \le t_m(d)=T$ such that a price type $k$ order is accepted
(rejected) if and only if $n\ge t_{k-1}(d)$ ($n< t_{k-1}(d)$); and 
for each $k$, $t_k(d)$ is decreasing in $d$.
}
\end{pro}

Throughout the paper, we use ``increasing'' and ``decreasing'' 
in the non-strict sense.


\medskip




\section{The Switch-Over Policy}
\label{sec:switch}

The structural result in Proposition \ref{pro:structure}
suggests a
threshold type policy  as follows.
For each type of order $i$, there is a critical state $(n_i,d_i)$
--- supply the order if and only when time has reached $n_i$ or
beyond and inventory is at least $d_i$. This policy, however, is
still difficult to analyze (in terms of deriving an explicit
objective function to be used for pricing optimization). What we do
below is to reduce this two-dimensional threshold policy to a single
dimension, in time only.
%
Specifically, we start with accepting only orders with the highest
price, until a time $t_1$, when we start to accept orders of the
top two prices, until a time $t_2$, when we start to accept orders
the top three prices, and so forth.
We call this a ``switch-over'' policy.

It is important to note that the switch-over policy
essentially follows the time epochs $t_\ell (d)$ ($\ell=1,\dots,m$) 
in Proposition \ref{pro:structure} but ignores their dependence 
on the inventory level $d$. Consequently, it is suboptimal
even in the context of Proposition \ref{pro:structure}.

In this section, we focus on the case of
a constant (unit) batch size, treating the general case of
random batch sizes in the
next section.


\subsection{Optimizing the Switch-Over Times}
\label{sec:unit}

We let the switch-over times be the
decision variables of an optimization problem, with the
objective to maximize the expected profit over the horizon -- the
same
objective as in the dynamic programming formulation.
(Note, however, since these time epochs
are chosen independently of the level of the
available inventory, the switch-over policy is in general
sub-optimal.)
To determine the best switch-over times,
\begin{eqnarray}
\label{switchtimes} 0=t_0\le  t_1 \le  t_2 \le \cdots \le t_{m-
1}\le t_m=T,
\end{eqnarray}
(only $t_1,...,t_{m-1}$ are decision variables), we consider a
continuous-time version of the original problem, with the order
streams following independent Poisson processes with rates
$\lam_i$, $i=1,...,m$, and associated with the price $p_i$, such
that
\begin{eqnarray}
\label{priceorder} p_1 \ge p_2 \ge \cdots \ge p_m.
\end{eqnarray}
Let $N_{1k}$ denote the total 
number of order arrivals of types $\{1,2, \cdots, k\}$ over
the time interval $(t_{k-1},t_k]$; $N_{1k}$ follows a Poisson
distribution with mean $(\g_1+\cdots+\g_k)(t_k-t_{k-1})$. 
Let
\begin{eqnarray}
\label{sprice} p_{1k}:= \frac{\lam_1p_1 +\cdots +\lam_k
p_k}{\lam_1+\cdots +\lam_k},
\end{eqnarray}
denote the average unit price of the orders accepted (by the
switch-over policy) over the time interval $(t_{k-1},t_k]$.
Then, we can write the objective function associated with the
switch-over policy as follows:
\begin{eqnarray}
\label{UD_revenue} \max &&p_{11}\ex\left[W\wedge N_1\right]+
p_{12}\ex[(W-N_{11})^+ \wedge N_{12}]
\nonumber\\
&&+ \cdots +
 p_{1m} \ex [(W-\sum_{k=1}^{m-1}N_{1k})^+ \wedge
N_{1m} ]
\nonumber\\
&=&p_{11}[W-\ex(W-N_{11})^+] +p_{12}[\ex (W-N_{11})^+
-\ex(W-N_1-N_{12})^+]
\nonumber\\
&&+ \cdots +
 p_{1m} [\ex (W-\sum_{k=1}^{m-1}N_{1k})^+
-\ex (W-\sum_{k=1}^{m}N_{1k})^+ ]
\nonumber\\
&=&p_{11}W- (p_{11}-p_{12})\ex (W-N_{11})^+ - \cdots -
 (p_{1,m-1}-p_{1m})\ex (W-\sum_{k=1}^{m-1}N_{1k})^+
\nonumber\\
&&- p_{1m}\ex (W-\sum_{k=1}^{m}N_{1k})^+ .
\end{eqnarray}
Denote
\begin{eqnarray}
\label{mean_variance}
\mu_{k}
&:=& \g_1t_k +
\g_2(t_k-t_1) + \cdots + \g_k (t_k-t_{k-1})\nonumber\\
&=& \g_1t_1 + (\g_1+\g_2)(t_2-t_1)+\cdots
+(\g_1+\cdots+ \g_k) (t_k-t_{k-1}) ,
\end{eqnarray}
which is the mean (as well as the variance) of $N_{11} + N_{12} +\cdots +
N_{1k}$.
Let  $N(\mu)$ denote a Poisson random variable with mean $\mu$.
 We can then turn the optimization problem in
(\ref{UD_revenue}) into the following equivalent form:
\begin{eqnarray}
\label{sswobj2}
\min_{0\le t_1\le\cdots\le t_{m-1} \le T}&&
(p_{11}-p_{12})\ex [ W-N(\mu_{1}) ]^+
+(p_{12} - p_{13})\ex [  W-N(\mu_{2})]^+ +\cdots \nonumber\\
&&+
(p_{1,m-1} - p_{1,m})\ex [  W-N(\mu_{m-1})]^+
+ p_{1m}\ex [  W-N(\mu_{m})]^+ .
\end{eqnarray}

Denote the distribution function of $N(\mu)$ as:
\begin{eqnarray}
\label{f}
F_n(\mu):=\pr [N(\mu)\le n]=\sum_{k=0}^n \frac{\mu^k}{k!}e^{-\mu};
\end{eqnarray}
and define a function:
\begin{eqnarray}
\label{h}
H(\mu):=\ex [W- N(\mu)]^+ =\sum_{k=0}^W (W-k)\frac{\mu^k}{k!}e^{-\mu}
=W F_W(\mu)-\mu F_{W-1}(\mu).
\end{eqnarray}
Clearly, $H(\mu)$ is decreasing and convex in $\mu$,
since $(x)^+$ is increasing and convex, and $N(\mu)$
is stochastically increasing and linear in $\mu$ (refer to \cite{shyao}).

From (\ref{h}), we can derive explicitly
$H'(\mu_\ell)$, which will become useful below. 
First, taking the derivative of $F_n (\mu)$ in
(\ref{f}), we have,
\begin{eqnarray}
\label{fder}
-f_n(\mu):=F'_n(\mu) =F_{n-1}(\mu)-F_n(\mu)=-\frac{\mu^n}{n!} e^{-\mu}.
\end{eqnarray}
Hence
\begin{eqnarray}
\label{hf}
H'(\mu) = -Wf_W(\mu )-F_{W-1}(\mu )+\mu f_{W-1} (\mu )
= -F_{W-1}(\mu ),
\end{eqnarray}
taking into account that
$f_W(\mu )=\frac{\mu}{W} f_{W-1} (\mu)$.
As a by-product, the above also leads to  
$$H''(\mu_\ell) = f_{W-1}(\mu_\ell) \ge 0,$$
confirming the convexity of $H(\mu)$.


Denote:
\begin{eqnarray}
\label{piy}
\pi_k:=p_{1k}-p_{1,k+1}, \qquad
y_k:=t_k-t_{k-1}, \qquad k=1,...,m;
\end{eqnarray}
 with $p_{1,m+1}:=0$, and
recall $t_0:=0$ and $t_m:=T$.
Then, the optimization problem in (\ref{sswobj2}) can be expressed
as
follows:
\begin{eqnarray}
\label{sobj}
\min_{(y_\ell)}\, \sum_{\ell =1}^{m} \pi_\ell
H(\mu_\ell) \qquad {\rm s.t.} \quad
 \sum_{\ell =1}^{m} y_\ell=T; \qquad y_\ell \ge 0, \quad \ell
=1,\dots ,m ;
\end{eqnarray}
where the new decision variables
are $(y_\ell)$; and,
following (\ref{mean_variance}) and (\ref{piy}),
\begin{eqnarray}
\label{meanvar}
\mu_{\ell}
=\Lam_1y_1 +\Lam_2y_2 +\cdots + \Lam_\ell y_\ell ,
\end{eqnarray}
with
\begin{eqnarray}
\label{lamsum}
\Lam_\ell:=\g_1+\cdots +\g_\ell.
\end{eqnarray}

\subsection{Solution Approach}
\label{sec:approach}

A shortfall of the above formulation is that the objective
function in (\ref{sobj}) is not separable in $(y_\ell)$ (although
the constraint is). If we make $(\mu_\ell)$ the decision variables
instead, then both the objective and the constraint will become
separable.
From (\ref{meanvar}),
we can solve for $(y_\ell)$:
\begin{eqnarray}
\label{yk} y_\ell = {\frac{\mu_\ell -\mu_{\ell-1}}
{\Lam_\ell }}, \qquad \ell = 1,...,m ;
\end{eqnarray}
with $\mu_0:=0$. We can then turn the optimization problem in
(\ref{sobj}) into one with $(\mu_\ell)$ as decision variables as
follows:
\begin{eqnarray}
\label{sobjsig}
\min_{(\mu_\ell)} && \sum_{\ell =1}^{m} \pi_\ell
H(\mu_\ell) \\
{\rm s.t.} &&
\sum_{\ell=1}^{m}
({1\over\Lam_\ell}-{1\over\Lam_{\ell+1}})\mu_\ell
 \le T , \label{equal}\\
&& 0\le \mu_1\le \cdots \le \mu_m; \label{increasing}
\end{eqnarray}
where
$\Lam_{m+1}^{-1}:=0$.
Also note that in (\ref{equal}) 
we have changed the original equality constraint ($=T$) 
 to an inequality constraint ($\le T$), 
taking into account that $H(\cdot)$ is a
decreasing function, and 
$\pi_\ell\ge 0$ as evident from (\ref{priceorder},
\ref{sprice},\ref{piy}).
Furthermore, as pointed out earlier, $H(\cdot)$ is convex;
hence, we now have a separable convex programming
problem. 
In addition, it has another appealing property
as revealed in the lemma below.

\begin{lem}
\label{l:monotone}
{\rm
The ratio of the coefficients in (\ref{equal}) to those in 
(\ref{sobjsig}), $({1\over\Lam_\ell}-{1\over\Lam_{\ell+1}})/\pi_\ell$,
is decreasing in $\ell$.
}
\end{lem}

{\startb Proof.}
We need to show:
\begin{eqnarray*}
{1\over \pi_{\ell-1}}( {1\over \Lam_{\ell-1}}- {1\over
\Lam_{\ell}} ) \ge {1\over \pi_{\ell}}( {1\over \Lam_{\ell}}-
{1\over \Lam_{\ell+1}} ),\qquad \ell=2,\dots, m.
\end{eqnarray*}
The above inequalities simplify to:
\begin{eqnarray}
\label{nonnegy1}
{\pi_\ell\over \pi_{\ell-1}}
{ \lam_\ell\over \lam_{\ell+1}} \ge { \Lam_{\ell-1}\over
\Lam_{\ell+1}} , \quad \ell=2,...,m-1;
\qquad{\rm and}\qquad
{\pi_m\over \pi_{m-1}}
{ \lam_m\over \Lam_{m-1}} \ge 1.
\end{eqnarray}
To verify these inequalities, write
$$\rho_\ell :=\sum_{k=1}^\ell \lam_kp_k
\nd
\pi_\ell=\frac{\rho_\ell}{\Lam_\ell}.$$
First consider $\ell\neq m$. Straightforward algebra yields:
\begin{eqnarray*}
{{\pi_\ell\Lam_{\ell+1}}\over {{\pi_{\ell-1}\Lam_{\ell-1}}}} &=&
{{\rho_\ell\Lam_{\ell+1} -\rho_{\ell+1}\Lam_{\ell}}\over
{\rho_{\ell-1}\Lam_{\ell} -\rho_{\ell}\Lam_{\ell-1}}}\\
&=& {{\rho_\ell\lam_{\ell+1}
-p_{\ell+1}\lam_{\ell+1}\Lam_{\ell}}\over
{\rho_{\ell-1}\lam_{\ell} -p_{\ell}\lam_\ell\Lam_{\ell-1}}}.
\end{eqnarray*}
We want the above to dominate $\lam_{\ell+1}/\lam_\ell$. This is
equivalent to
$$\rho_\ell -p_{\ell+1}\Lam_{\ell}
\ge \rho_{\ell-1} -p_{\ell}\Lam_{\ell-1},$$ or,
\begin{eqnarray*}
p_\ell\lam_\ell \ge
p_{\ell+1}\Lam_{\ell} -p_{\ell}\Lam_{\ell-1}
= p_{\ell}\lam_{\ell} -(p_{\ell}-p_{\ell+1})\Lam_{\ell},
\end{eqnarray*}
which obviously holds, since $p_\ell \ge p_{\ell+1}$.
Next, we show the last inequality in (\ref{nonnegy1}).
Since
$$\pi_m=\frac{\rho_m}{\Lam_m},\qquad
\pi_{m-1}=\frac{\rho_{m-1}}{\Lam_{m-1}}-\frac{\rho_m}{\Lam_m},$$
we have
$$\frac{\pi_m}{\pi_{m-1}}
=\frac{\rho_m \Lam_{m-1}}{\rho_{m-1}\Lam_{m}-\rho_m\Lam_{m-1}}.$$
We want the above to dominate
$\frac{\Lam_{m-1}}{\lam_m}$; and this simplifies to
$$\rho_m\lam_m \ge \rho_{m-1}\Lam_m -\rho_{m}\Lam_{m-1},$$
or $\rho_m\Lam_m \ge \rho_{m-1}\Lam_m$, i.e.,
$\rho_m \ge \rho_{m-1}$, which certainly holds.
\hfill$\Box$

\medskip

The above lemma suggests an algorithm as follows.
Ignoring the constraint in (\ref{increasing}), and with $\eta \ge 0$
denoting the Lagrangian multiplier associated with the constraint
in (\ref{equal}), we have 
the following optimality equations:
\begin{eqnarray}
\label{nonlinear}
 F_{W-1}(\mu_\ell)=\frac{\eta}{\pi_\ell}
\Big( {1\over {\Lam_{\ell}}}- {1\over {\Lam_{\ell+1}}}\Big),
\quad \ell=1,...,m.
\end{eqnarray}
(Recall from (\ref{hf}), we have $F_{W-1}(\mu)=-H'(\mu)$.)
From (\ref{fder}), we know $F_{W-1}(\mu)$ is decreasing 
in $\mu$; and from Lemma \ref{l:monotone}, we know the right hand
sides of the equations above are also decreasing in $\ell$. Hence,
the solution to these optimality equations, $\mu_\ell$, must be increasing in $\ell$. 

To solve these equations, we can start with an
 $\eta$ value sufficiently close to zero,
so that each of the $m$ equations in 
(\ref{nonlinear}) has a solution.
(Note that $F_{W-1}(\mu)\downarrow 0$ when $\mu\to\infty$.)
This may very well violate the constraint in (\ref{equal}), 
as the resource is priced too low.
Hence, we will gradually increase $\eta$, to bring down the 
left hand side of the constraint in (\ref{equal}), until it  
is satisfied as an equality. 
(Note that since $F_{W-1}(\mu)$ is decreasing in $\mu$, 
increasing $\eta$ will decrease the value of $\mu_\ell$'s 
that solve the equation in (\ref{nonlinear}).)
As we increase $\eta$, it can happen that for some $k$, even with 
$\mu_k=0$, we still have
\begin{eqnarray}
\label{rk}
R_k:=\frac{ F_{W-1}(0)}
{\frac{1}{\pi_k}\Big({1\over\Lam_k}-{1\over\Lam_{k+1}}\Big)} <\eta.
\end{eqnarray}
Then, we set $\mu_k=0$, as any $\mu_k>0$ will only 
lead to $F_{W-1}(\mu_k)<F_{W-1}(0)$.
In this case we also set 
$\mu_j=0$ for all $j <k$ as
well, since $R_j\le R_k$ following Lemma \ref{l:monotone}.
Furthermore, $\mu_1=\cdots =\mu_k=0$ will remain at zero as 
$\eta$ further increases. 
(As $\eta$ increases, more $\mu_\ell$'s may become zero, for $\ell >k$.)

\begin{pro}
\label{pro:singleunit}
{\rm
(i) The optimal switch-over times 
can be obtained from 
solving the set of single-variable
equations in (\ref{nonlinear}) along with an increasing sequence of 
$\eta$ (the Lagrangian multiplier) 
until the constraint in (\ref{equal}) becomes binding.

\noindent
(ii) At optimality, for any price class $k$
we have $\mu_k=0$ if and only if
$R_k <\eta$ in (\ref{rk});
and if $\mu_k=0$ for some $k$, we must have 
$\mu_j=0$ for all $j\le k$.
}
\end{pro}

{\startb Proof.}
What is stated in (i) and (ii) is a summary of the algorithm
and the solution it generates.
 Hence, what needs to be argued is the optimality
of the solution. To do so,
in addition to the Lagrangian multiplier $\eta \ge 0$, 
let $\nu_\ell\ge 0$ be the Lagrangian multiplier associated with 
the constraint $\mu_\ell\ge \mu_{\ell-1}$, for $\ell =1,\dots, m$
(with $\mu_0:=0$). 
In addition, denote $\nu_{m+1}:=0$.
Then, optimality is reached if and only if $(\mu_1, \dots, \mu_m)$
and all the multipliers satisfy the following equations:  
\begin{eqnarray}
&&\pi_\ell F_{W-1}(\mu_\ell)=\eta 
\Big( {1\over {\Lam_{\ell}}}- {1\over {\Lam_{\ell+1}}}\Big) -\nu_\ell +\nu_{\ell+1},
\quad \ell=1,...,m 
 ; \label{opt1} \\
&&\eta \Big[ \sum_{\ell=1}^{m} 
({1\over\Lam_\ell}-{1\over\Lam_{\ell+1}})\mu_\ell
 -T\Big] =0 ; \label{opt2}\\
&& \nu_\ell ( \mu_\ell -\mu_{\ell-1})=0, 
\quad \ell=1,...,m ; \label{opt3} 
\end{eqnarray}
and the constraints in (\ref{equal}) and (\ref{increasing}) are satisfied.

As described above, when  the algorithm ends, 
it generates a solution that
takes the following form:
\begin{eqnarray}
\label{optsolution}
0:=\mu_0=\mu_1 =\cdots = \mu_{k}< \mu_{k+1}\le\mu_{k+2}\le \cdots \le \mu_m ,
\end{eqnarray}
for some $k \in\{1,\dots,m\}$, along with an $\eta \ge 0$, while 
(\ref{equal}) is satisfied as an equality -- and hence (\ref{opt2}) holds. 
In addition, (\ref{increasing})  follows from (\ref{optsolution}).
Hence, what remains is to verify (\ref{opt1}) and (\ref{opt3}).
These can be satisfied by letting
\begin{eqnarray}
\nu_\ell&=&0, \qquad \ell=k+1,\dots, m; \label{nu0}\\
\nu_k&=&\eta\Big( {1\over {\Lam_{k}}}- {1\over {\Lam_{k+1}}}\Big)
 -\pi_k F_{W-1}(0); \label{nu1}\\
\nu_j&=& \nu_{j+1}+\eta\Big( {1\over {\Lam_{j}}}- {1\over {\Lam_{j+1}}}\Big)
 -\pi_j F_{W-1}(0), \qquad j=k-1, \dots, 1 . \label{nu2}
\end{eqnarray}
First, the $\nu$'s above are all non-negative:
$\nu_k>0$ follows from (\ref{rk}); and for $j<k$, $\nu_j>0$ follows recursively
from (\ref{nu2}), 
as well as from (\ref{rk}), taking into account $R_j\le R_k< \eta$ for $j<k$.
Second, the above $\nu$'s, along with the $\mu$'s in (\ref{optsolution}),
satisfy (\ref{opt3}). 
Finally, note that the zero $\nu_\ell$'s ($\ell =k+1,\dots, m$) 
in (\ref{nu0}) reduce (\ref{opt1}) to 
the equations (with the same $\ell$ indices)
in (\ref{nonlinear}) used by the algorithm; and
substituting (\ref{nu1}) and (\ref{nu2}) into (\ref{opt1}) recovers
$\mu_j=0$ obtained by the algorithm for $j\le k$. 
 \hfill$\Box$

\section{The Switch-Over Policy: Random Batch Size}
\label{sec:batch}

We now extend the switch-over policy of the last section to allow
random batch sizes. In this case, while the structural properties in 
Proposition \ref{pro:structure} do not apply, 
the asymptotic optimality results in the next section does provide some
justification to the switch-over policy.

\subsection{Homogeneous Batches}
\label{sec:independent}

To facilitate the
derivation, we start with assuming
that all orders independently follow an identical batch-size
distribution that is independent of the prices, deferring
the case of price-dependent batch sizes to the next subsection.
Specifically, the batch size distribution is denoted
by $q_j =\pr\{Q=j\}$, for $j=1,...,W$.

With the initial inventory of $W$ units,
we apply the switch-over policy in $[0,t_1]$.
The cumulative revenue collected by time
$t_1$ is:
$$ p_1 \ex [W-{\bf z}^T M^{N_1}{\bf w}],$$
where,
$$ {\bf z}^T:=(0,0,...,0,1), \qquad {\bf w}^T:= (0,1,2,...,W);$$
$N_1:=N_1(t_1)$ follows the Poisson distribution
with mean $\g_1 t_1$, and $M$ denotes the 
following probability transition matrix:
\begin{eqnarray}
\label{TransitionMatrix} M=\left( \begin{array}{ccccc} 1 & 0 & 0&
... & 0 \\ q_1& 1-q_1& 0&...&
0\\q_2&q_1&1-q_1-q_2&...& 0 \\
&&&...&\\q_W&q_{W-1}&q_{W-2}&...&1-q_1-...-q_W \end{array} \right)
.
\end{eqnarray}
That is, the number of remaining units in the system,
embedded at the arrival epochs of orders of price $p_1$, is
a Markov chain with the probability transition matrix $M$.
Note that the dimension of $M$ is $W+1$, corresponding to the
dimension of the state space of the Markov chain, $\{0,1,...,W\}$.

Recall that the generating function of a Poisson variable $N$ with
mean $\mu$ is:
$$\ex [z^N] = e^{-\mu} e^{\mu z} .$$
Hence, we have
\begin{eqnarray*}
p_1 \ex [W-{\bf z}^T M^{N_1}{\bf w}]
= p_1[W-e^{-\mu_1}{\bf z}^Te^{\mu_1  M}{\bf w}].
\end{eqnarray*}
Here, $\mu_1$ follows the notation defined in
(\ref{meanvar}); and the matrix exponent is defined as
\begin{eqnarray*}
e^A=\sum_{j=0}^{\infty} \frac{A^k}{k!},
\end{eqnarray*}
for any matrix $A$.

Similarly, for the next interval $(t_1,t_2]$, when the acceptance
includes both price classes $p_1$ and $p_2$, the expected revenue
is (using the notation of the last section):
\begin{eqnarray*}
&&p_{12}\ex[{\bf z}^T M^{N_1} {\bf w}
-{\bf z}^T M^{N_1+N_{12}}{\bf w}] \\
&=&p_{12}[{\bf z}^T(e^{-\mu_1}e^{\mu_1 M}-
e^{-\mu_2}e^{\mu_2 M}){\bf w}] .
\end{eqnarray*}

Comparing the above with the objective function in
(\ref{UD_revenue}), we know the only difference is that the terms
there,
$$\ex(W-N_1)^+ \nd \ex(W-N_1-N_{12})^+$$
are now replaced by
$${\bf z}^T e^{-\mu_1}e^{\mu_1 M} {\bf w} \nd
{\bf z}^T e^{-\mu_2}e^{\mu_2 M} {\bf w} .$$
With this in mind, we
further denote
\begin{eqnarray}
\label{g}
G(\mu):={\bf z}^T
e^{-\mu}e^{\mu M}  {\bf w},
\end{eqnarray}
which is analogous to $H(\mu )$ of the last section.
The derivatives of $G(\mu)$ can be derived as follows:
\begin{eqnarray}
\label{gderivatives}
G'(\mu)= -{\bf z}^T e^{-\mu} (I-M) e^{\mu M}  {\bf w},
\qquad G''(\mu)= {\bf z}^T
e^{-\mu} (I-M)^2 e^{\mu M}  {\bf w}.
\end{eqnarray}
Note that the spectral radius of the
matrix
$M$ is unity. Hence, $G'(\mu)\le 0$; i.e., $G(\mu)$ is a
decreasing function.
Furthermore, $G(\mu)$ is a convex function,
since $G''(\mu)\ge 0$.

We can formulate the optimization problem as follows,
with $\mu_\ell$, $\ell =1,...,m$, as decision variables:
\begin{eqnarray}
\label{sobjmu}
\min_{(\mu_\ell)}\, \sum_{\ell =1}^{m} \pi_\ell G(\mu_\ell);
\qquad {\rm s.t.} \; \sum_{\ell=1}^{m} (
{1\over\Lam_\ell}-{1\over\Lam_{\ell+1}})\mu_\ell
\le T , 
\qquad 0\le \mu_1\le\cdots\le\mu_m .
\end{eqnarray}
Here, same as in (\ref{yk}),
\begin{eqnarray*}
y_\ell:=
t_\ell -t_{\ell -1} = {{\mu_\ell -\mu_{\ell-1}}\over {\Lam_\ell
}}, \qquad \ell = 1,...,m ;
\end{eqnarray*}
with $\mu_0:=0$.

To solve the optimization problem above, we 
can follow the same approach as in the last section;
specifically, we can solve
the following optimality equations via gradually increasing
the Lagrangian multiplier $\eta$
until the resource 
constraint becomes binding:  
 \begin{eqnarray}
 {\bf z}^T e^{-\mu_\ell} (I-M) e^{\mu_\ell M}  {\bf w}
={\eta\over\pi_\ell}
 ( {1\over {\Lam_{\ell}}}- {1\over {\Lam_{\ell+1}}}),
\qquad \ell=1,...,m .
\end{eqnarray}
Note that both Lemma \ref{l:monotone} and
Proposition \ref{pro:singleunit} still apply here.

\subsection{Price-Dependent Batches}
\label{sec:dependent}

Now, suppose each price $p_i$, $i=1,\dots,m$, is associated with a
batch size $Q_i$, with the distribution $\pr (Q_i = j)=\th_{ij}$,
for $j=1,...,W$. The batch sizes are i.i.d.\ among orders of the
same price, and independent among orders of different prices.
Here, instead of a single probability transition matrix $M$, we
have
$m$ such matrices, one for each price $p_i$:
\begin{eqnarray}
\label{TransitionMatrixPrice} M_i=\left( \begin{array}{ccccc} 1 &
0 & 0& ... & 0 \\ \th_{i1}& 1-\th_{i1}& 0&...&
0\\\th_{i2}&\th_{i1}&1-\th_{i1}-\th_{i2}&...& 0 \\
&&&...&\\
\th_{iW}&\th_{i,W-1}&\th_{i,W-2}&...&1-\th_{i1}-...-\th_{iW}
\end{array} \right) .
\end{eqnarray}

Consider the second time interval $(t_1,t_2]$. Since orders of
both prices $p_1$ and $p_2$ are accepted, the two Poisson streams
will be combined, resulting in the following transition matrix
$$\Gamma_{2}:=\frac{1}{\Lam_{2}} (\g_1M_1+
\g_2M_2).$$
Analogously, denote $\Gamma_1 := M_1$. The expected
revenue over this time interval is
\begin{eqnarray*}
&&p_{12}\ex[{\bf z}^T M_1^{N_1} {\bf w}
-{\bf z}^T M_1^{N_1}\Gamma_{2}^{N_{12}} {\bf w}] \\
&=&p_{12}[{\bf z}^T(e^{-\mu_1}e^{\mu_1 \Gamma_1}-
e^{-\mu_2}e^{\mu_1 \Gamma_1}e^{(\mu_2-\mu_1)\Gamma_2}){\bf w}] .
\end{eqnarray*}

In general, denote
$$\Gamma_\ell:=\sum_{j=1}^\ell\frac{\g_j}{\Lam_\ell}M_j.$$
Replace $G(\mu_\ell)$ in the earlier special case of
independent batch sizes by the following:
\begin{eqnarray}
\label{gg}
g_\ell (\by_\ell):=e^{-\mu_\ell}\exp(\sum_{k=1}^\ell
\Lam_ky_k\Gamma_k )
\nd
G_\ell (\by_\ell):= {\bf z}^T g_\ell (\by_\ell){\bf w}.
\end{eqnarray}
(Notice that from (\ref{meanvar}),
we have $\mu_\ell -\mu_{\ell-1}=\Lam_\ell y_\ell$.)

The optimization problem can now be expressed as follows,
with $y_\ell$, $\ell =1,...,m$, as decision variables:
\begin{eqnarray}
\label{sobjy}
\min_y\, \sum_{\ell =1}^{m} \pi_\ell G_\ell(\by_\ell);
\qquad {\rm s.t.} \;
 \sum_{\ell =1}^{m} y_\ell \le T;
\qquad y_\ell \ge 0, \quad \ell =1,...,m .
\end{eqnarray}
From (\ref{gg}), we know that
$g_\ell(\by_\ell)$ is
convex in $\by$; hence,  the problem in (\ref{sobjy}) is a
convex program (albeit no longer separable) 
and as such can be solved by
standard algorithms.

\subsection{A Numerical Example}
\label{sec:numerical}


\begin{exm}
{\rm
Consider four price classes:
$$ p_1=1,\quad  p_2=0.8,\quad  p_3=0.65,\quad p_4 =0.45,$$
with Poisson arrivals at the following rates:
$$ \g_1=0.2,\quad \g_2=0.3,\quad \g_3=0.1, \quad \g_4=0.4.$$
The batch size $Q$ is homogeneous among the four classes,
following
a discretized exponential distribution
with mean $12$.
(Specifically, $\pr (Q=n)=e^{\g (n+1)}-e^{\g n}$, for
$n=0,1,2...$, and $1/\g =12$.)
 Let $T=20$,
and let $W$ vary from $1$ to $100$.

We compare the performance of the switch-over policy against the
optimal policy (from DP) and a first-come-first-served (FCFS)
policy (i.e. supply any order on arrival as long as there are
units available). The objective values corresponding to
 the three policies are
plotted in Figure 1.  The relative errors of the switch-over
policy and the FCFS policy
with respect to the optimal policies are plotted in Figure 2. The
performance of the switch-over policy is remarkably close to the
optimal policy, while the performance of the FCFS policy, as
expected, deteriorates quickly as $W$ increases.
\hfill$\Box$
}
\end{exm}


\resizebox{140mm}{!}{\includegraphics*[0in,0in][12in,8in]{figure1.eps}}

\resizebox{140mm}{!}{\includegraphics*[-0.1in,0in][12in,8in]{figure2.eps}}

\section{Asymptotic Optimality of the Switch-Over Policy}
\label{sec:asymptotic}

In this section, we establish the fact that the switch-over policy
is asymptotically optimal, in the sense that as $W$ increases to
infinity, the relative error between the switch-over policy and
the optimal policy (via dynamic programming) will go to zero. In
fact, we show the absolute error is no more than $O(\sqrt{W})$;
hence, the rate of convergence is $O(\sqrt{W})$. This type of
asymptotic optimality is in the same spirit as other recent studies, e.g.,
\cite{mz}. Similar results have been established for various revenue
management problems, see,
e.g.\ \cite{cooper,gallegovanryzin,tallurivanryzin}.
The result here applies to
both models in the last two sections; for ease of exposition, we
shall focus on the model in \S\ref{sec:switch}.

Intuitively, when the available inventory
$W\to\infty$, we should expect the switch-over policy to be
optimal in the following two cases: (a) if the decision horizon
$T$ remains a constant or is of lower-order than $W$, then
supplying all orders (i.e., first-come-first-served) would be
optimal;
and (b) if $T$ is of higher order than $W$,
then supplying the
highest bid only (orders priced at $p_1$) should
be optimal.
In both cases the policies can be represented as special cases
of the switch-over policy.
Below, we will formally
confirm the intuition behind
these two special cases, and also prove the result in
the non-trivial case
in which $T\to\infty$ in the same order as $W$.

We start from an upper-bound on the performance of the optimal
policy.

\begin{lem}
\label{lem:upper}
{\rm
The objective value under the
optimal policy (as derived from the dynamic programming),
denoted $V^*$, satisfies the following inequality:
\begin{eqnarray}
\label{optup}
V^*\le (\lam_1p_1+\cdots +\lam_kp_k)T+\lam_{k+1} p_{k+1} t,
\end{eqnarray}
where $k=1,...,m-1$ and $t\in [0,T)$ are such that
$\Lam_k T\le W$, $\Lam_{k+1} T >W$, and
\begin{eqnarray}
\label{wk}
\Lam_{k} T +\lam_{k+1} t =W.
\end{eqnarray}
If $\Lam_1 T> W$, then $V^*\le p_1 W$.
If $\Lam_m T\le W$, then $V^*\le p_{1m}\Lam_m T$.
}
\end{lem}

{\startb Proof.}
Note that $V^*$ is maximal among all non-anticipative
policies; in particular, at each time $t$ all future order
arrivals
and their prices are only known in {\it distribution}.
Now, suppose at time $t=0$ we know all arrivals over $(0,T]$
and their prices {\it deterministically}.
Specifically, suppose over $(0.T]$, there are $N_1$ orders of
price $p_1$,
$N_2$ orders of price $p_2$, ..., and $N_m$ orders of price $p_m$.
Then, we will obviously use the $W$ available units to
first supply the $N_1$ orders of price $p_1$; and then,
if there is anything left, supply the
$N_2$ orders of price $p_2$; and so forth.
Clearly, no non-anticipative policy can do better than this
one, which yields an objective value of
\begin{eqnarray*}
&&p_1[W\wedge N_1]+
p_{2}[(W-N_{1})^+ \wedge N_{2}]
+ \cdots + p_{m} [(W-\sum_{k=1}^{m-1}N_{k})^+ \wedge N_{m} ]
\nonumber\\
&=& p_{1}W- (p_{1}-p_{2})(W-N_1)^+ - \cdots -
 (p_{m-1}-p_{m})(W-\sum_{k=1}^{m-1}N_{k})^+
- p_{m}(W-\sum_{k=1}^{m}N_{k})^+ .
\end{eqnarray*}
Taking expectation on the right hand side above
and passing the expectation into $(\cdot )^+$ further
increases its value (since the
function $(x)^+$ is convex).
Hence, we have
\begin{eqnarray*}
V^* &\le&
p_{1}W- (p_{1}-p_{2})(W-\lam_1T)^+ - \cdots -
 (p_{m-1}-p_{m}) (W-\sum_{k=1}^{m-1}\lam_{k} T)^+
\nonumber\\
&& - p_{m}(W-\sum_{k=1}^{m-1}\lam_{k}T)^+
\nonumber\\
&=& p_{1}W- (p_{1}-p_{2})(W-\lam_1 T) - \cdots -
 (p_{k}-p_{k+1}) (W- \lam_1 T -\cdots -\lam_{k}T)
\nonumber\\
&=& p_{k+1}W +(p_{1}-p_{k+1})\lam_1 T+(p_{2}-p_{k+1})\lam_2 T +
 \cdots + (p_{k}-p_{k+1}) \lam_k T
\nonumber\\
&=&
(\lam_1p_1+\cdots +\lam_kp_k)T+\lam_{k+1} p_{k+1} t ,
\end{eqnarray*}
where the last two equalities follow from
the relation in (\ref{wk}).

Following the same reasoning also leads to the bounds
in the two cases when $\Lam_1 T> W$ or
$\Lam_m T\le W$.
\hfill$\Box$

\medskip

Next, we derive a lower bound on the switch-over policy.

\begin{lem}
\label{lem:lower}
{\rm
Let $V^{\rm sw}$ denote the objective value
under the best switch-over policy. Then,
letting $k$ and $t$ be defined as in
Lemma \ref{lem:upper}, we have
\begin{eqnarray}
\label{vlower}
V^{\rm sw}
&\ge& p_{1k}W -(p_{1k}-p_{1,k+1})\ex [W-N(\Lam_k (T-t))]^+
\nonumber\\
&& -p_{1,k+1}\ex [W-N(\Lam_k T+\lam_{k+1}t)]^+ ;
\end{eqnarray}
and
\begin{eqnarray*}
V^{\rm sw}\ge p_1 \ex[N(W)\wedge W] &
{\rm if} & \Lam_1 T> W, \\
V^{\rm sw}\ge p_{1m} \ex[N(\Lam_{m} T)\wedge W] &
{\rm if} & \Lam_m T\le W .
\end{eqnarray*}
}
\end{lem}

{\startb Proof.}
Clearly,
a feasible switch-over policy, realizable by
setting $t_1=\cdots =t_k=0$ and $t_{k+1}=T-t$,
is to accept the top $k$ price classes throughout the horizon
$[0,T]$ and
the class $k+1$ over $[T-t,T]$,  the last $t$ time units.
Hence,
\begin{eqnarray*}
V^{\rm sw}
&\ge& p_{1k}\ex [N(\Lam_k (T-t))\wedge W]
+p_{1,k+1} \ex \{N(\Lam_{k+1}t)\wedge [W-N(\Lam_k (T-t))]^+\}
\nonumber\\
&=& p_{1k}W -p_{1k}\ex [W-N(\Lam_k (T-t))]^+
\nonumber\\
&& +p_{1,k+1} \ex [W-N(\Lam_k (T-t))]^+
-p_{1,k+1}\ex [W-N(\Lam_k (T-t))-N(\Lam_{k+1}t)]^+
\nonumber\\
&=& p_{1k}W -(p_{1k}-p_{1,k+1})\ex [W-N(\Lam_k (T-t))]^+
-p_{1,k+1}\ex [W-N(\Lam_k T+\lam_{k+1}t)]^+ \nonumber\\
\end{eqnarray*}
(Recall, $p_{1k}$ is the average price among classes 1 through
$k$,
defined in (\ref{sprice}), and $p_{1,k+1}$ is similarly defined.)
If $\Lam_1 T> W$, then serving class 1 only throughout $[0,T]$
results in
$$V^{\rm sw}\ge p_1 \ex[N(\Lam_1 T)\wedge W]\ge p_1 \ex[N(W)\wedge
W].$$
If $\Lam_m T\le W$, then serving all $m$ classes throughout
$[0,T]$ leads to
$V^{\rm sw}\ge p_{1m} \ex[N(\Lam_{m} T)\wedge W].$
\hfill$\Box$

\medskip

We shall also need the following result:
For the Poisson variate with mean $a$, denoted $N(a)$,
\begin{eqnarray}
\label{poissonH}
\ex [a-N(a)]^+ \sim \sqrt{\frac{a}{2\pi}},
\qquad a\to\infty;
\end{eqnarray}
which can be directly verified via the Poisson
distribution (along with the Stirling formula), or
via the normal approximation of the Poisson distribution.

\begin{pro}
{\rm The switch-over policy is asymptotically optimal, in the
sense that $V^{\rm sw}/V^*\rightarrow 1$
when $W\to \infty$.
}
\end{pro}

{\startb Proof.}
To start with, suppose while $W\to \infty$, $T$ remains a
constant.
This corresponds to the case of $\Lam_m T < W$
in Lemmas \ref{lem:upper} and \ref{lem:lower}, and we have
\begin{eqnarray*}
p_{1m}\Lam_m T \ge V^*\ge V^{\rm SW} &\ge&
p_{1m}\ex [N(\Lam_m T)\wedge W] \\
&\to& p_{1m}\ex [N(\Lam_m T)
=p_{1m}\Lam_m T ,
\end{eqnarray*}
 where the limit follows from monotone
convergence as $W\to\infty$.
Hence, in this case we have
$V^*= V^{\rm SW}=p_{1m}\Lam_m T$.

Now, suppose $T\to\infty$; and without loss of generality, suppose
as a function of $W$, $T(W)$ satisfies the following:
$$0\le c:= \liminf\frac{T(W)}{W}\le \infty.$$
That is, we allow $T$ to be of lower
or higher order than $W$, as well as of equal order to $W$.
First, suppose
$\Lam_1^{-1}\le c\le \Lam_m^{-1}$.
Then, we can assume the relation between $W$ and $T$ in (\ref{wk})
to hold,
for some $k$.
Note that the upper bound in (\ref{optup}) can be written as:
\begin{eqnarray*}
&&(\lam_1p_1+\cdots +\lam_kp_k)T+\lam_{k+1} p_{k+1} t \\
&=&\Lam_kp_{1k}T+\lam_{k+1} p_{k+1} t \\
&=&p_{1k}W- (p_{1k}-
p_{k+1})\lam_{k+1} t.
\end{eqnarray*}
Hence, combining (\ref{optup}) and (\ref{vlower}), we have
\begin{eqnarray}
\label{errorbound}
0&\le& V^*- V^{\rm sw}
\nonumber\\
&\le&-(p_{1k}-p_{k+1})\lam_{k+1}t+(p_{1k}-p_{1,k+1})
\ex [W-N(\Lam_k (T-t))]^+
\nonumber\\
&&+p_{1,k+1}\ex [W-N(\Lam_k T+\lam_{k+1}t)]^+ .
\end{eqnarray}
Since
$$p_{1k}(\Lam_{k+1}-\lam_{k+1})
=\lam_1p_1+\cdots +\lam_kp_k
= p_{1,k+1}\Lam_{k+1}- p_{k+1}\lam_{k+1},$$
we can write
$$(p_{1k}-p_{k+1})\lam_{k+1}t =
(p_{1k}-p_{1,k+1})[W-\Lam_k (T-t)]. $$
Furthermore,
\begin{eqnarray*}
\ex [W-N(\Lam_k (T-t))]^+
&=&\ex [W-\Lam_k (T-t)+\Lam_k (T-t)-N(\Lam_k (T-t))]^+ \\
&\le & W-\Lam_k (T-t)+\ex [\Lam_k (T-t)-N(\Lam_k (T-t))]^+,
\end{eqnarray*}
where the inequality follows from $W\ge\Lam_k (T-t)$.
Hence,
the first two terms on the right hand side of (\ref{errorbound})
can be combined to yield the following:
%
\begin{eqnarray*}
0&\le&  V^*- V^{\rm sw}\\
&\le& (p_{1k}-p_{1,k+1})
\ex [\Lam_k(T-t)-N(\Lam_k (T-t))]^+]+ p_{1, k+1} \ex[W-N(W)]^+ .
\end{eqnarray*}
From (\ref{poissonH}), we know the two terms on the right side
above are of order $O(\sqrt{T})$ and $O(\sqrt{W})$, respectively.
On the other hand, $V^*$ is clearly of order $O({W})$ ---
both the upper and lower bounds
in Lemmas \ref{lem:upper} and \ref{lem:lower} are of order $O(W)$.
Hence, dividing both sides above by $V^*$ and letting $W\to
\infty$
results in $V^{SW}/V^* \to 1$.

Next, consider
the case of $c>\Lam_1^{-1}$,  which corresponds to the case of
$\Lam_1 T>W$.
From Lemmas \ref{lem:upper} and \ref{lem:lower}, we have
\begin{eqnarray*}
0\le  V^*- V^{\rm sw}
&\le& p_1W-p_1\ex[W\wedge N(W)]^+ \\
&=& p_1\ex[W-N(W)]^+ \sim O(\sqrt{W}).
\end{eqnarray*}
In this case, $V^*$ is still of order $O({W})$; and hence, $V^{\rm
SW}/V^* \to 1$.
Finally, in the case of $c<\Lam_m^{-1}$,
which corresponds to $\Lam_m T< W$,
we have
\begin{eqnarray*}
0\le  V^*- V^{\rm sw}
&\le& p_{1m}\Lam_m T-p_{1m}\ex[W\wedge N(\Lam_m T)]^+ \\
&\le& p_{1m}\Lam_m T-
p_{1m}\ex[\Lam_m T\wedge N(\Lam_m T)]^+ \\
&=& p_{1m}\ex[\Lam_m T - N(\Lam_m T)]^+ \sim O(\sqrt{T}).
\end{eqnarray*}
In this case, $V^*$ is of order $O({T})$;
hence, $V^{\rm SW}/V^* \to 1$ follows when $T\to\infty$.
\hfill$\Box$

\begin{exm}
\label{asymexm1}
{\rm
We continue with the example in
\S\ref{sec:numerical},
but suppose
the batch size $Q$ follows negative binomial distributions:
$$\pr[Q=k]=\binom{k+r-1}{r-1}p^r(1-p)^{k}, \qquad  k=0, 1, 2,
\cdots;$$
and hence,
$$ \ex[Q]=\frac{r(1-p)}{p},\qquad \var[Q]= \frac{r(1-p)}{p^2}. $$
We choose $(r,p)$ to be $(4,0.33)$ and $(8, 0.5)$.
While both distributions have the same mean $8$, their
coefficients of variation are $3$ and $2$, respectively.

In Table \ref{tblbatch}, we compare the performance between the
optimal policy and the switch-over policy, with the switch-over
times optimized (the column under ``Switch''). Fixing $T=20$, we
increase the value of $W$.
The relative error between the two policies first has a
slight increase, and then quickly decreases as $W$ becomes
large (8-10 times the value of $T$).
In addition, we
list in the table the performance of a switch-over policy with
equally spaced switch-over times (the column under ``Equal''),
 i.e., $y_i=T/4=5$ for all $i$.
The results indicate that the performance of this policy
deteriorates rather severely as $W$ increases.
That is, it is crucial to optimize the switch-over times.

Next, in Table \ref{tblbatch_asm}, we increase $T$ simultaneously
with $W$.
The relative error of the switch-over policy
appears to decrease monotonically.
\hfill$\Box$
}
\end{exm}

\medskip

\begin{table}
\centering
\begin{tabular}{||c|c|c|cc|cc||}\hline 
$W$& $(r,p)$ & Optimal  &  Switch & \% off & Equal & \% off \\
\hline
20
   & (4,0.33) &
17.59  & 17.47  & 0.67\% & 17.34 & 1.37\% \\
  & (8,0.5) &
17.68  & 17.50  & 0.99\% & 17.28 & 2.27\% \\
 \hline
40
   & (4,0.33) &
34.39  & 34.16  & 0.67\% & 33.99 & 1.18\% \\
   & (8,0.5) &
34.60  & 34.26  & 0.97\% & 34.14 & 1.34\% \\
 \hline
60
  & (4,0.33) &
50.08  & 49.62  & 0.92\% & 48.74 & 2.69\% \\
   & (8,0.5) &
50.39  & 49.84  & 1.09\% & 49.14 & 2.48\% \\
\hline 160
  & (4,0.33) &101.43&  100.87&  0.55\% & 72.85& 28.18\%
 \\
   & (8,0.5) &102.38&   101.84& 0.52\% & 72.72&  28.97\%
\\
\hline 180
  & (4,0.33) &104.54 &104.37& 0.16\% &72.69& 30.47\%
 \\
   & (8,0.5) &105.48& 105.31&  0.16\% &73.14 & 30.66\%
\\
   \hline 200
  & (4,0.33) &
105.74  & 105.69 & 0.05\% &72.73  & 31.20\% \\
   & (8,0.5) &
105.78  & 105.76  & 0.01\% & 72.56 & 30.82\% \\
\hline \hline
\end{tabular}
\caption{\label{tblbatch} Performance of the policies
as $W$ increases (with fixed $T$).}
\end{table}

\medskip

\begin{table}
\centering
\begin{tabular}{||c|c|c|c|c||c|c|c|c|c||}\hline 
$W, T$& $(r,p)$ & Optimal  &  Switch  & \% off &$W, T$& $(r,p)$
& Optimal  &  Switch  &\% off \\
\hline 20, 20
& (4,0.33) &17.59  & 17.47
&0.67\%   & 40, 40 & (4,0.33) & 36.74  & 36.55  & 0.52\%\\ &
(8,0.5) &
17.68  & 17.50  & 0.99\% & & (8,0.5) &
36.87  & 36.60 & 0.73\%   \\
 \hline
60, 60
  & (4,0.33) &
55.62  & 55.38  & 0.43\% &80, 80 & (4,0.33) &
75.11 & 74.95  & 0.21\% \\
   & (8,0.5) &
55.71 & 55.41  & 0.54\% & & (8,0.5) & 75.14 & 74.97  & 0.23\% \\
\hline
\end{tabular}
\caption{\label{tblbatch_asm}  Performance of the switch-over
policy with
simultaneous increase of both $W$ and $T$.}
\end{table}

\section{Pricing Models}
\label{sec:pricing}

As mentioned in the introductory section, an important motivation
for us to study the switch-over policy is to use it as a means to
solve optimal pricing problems. (In this sense, the prices derived
below are only optimal with respect to the switch-over policy.)
Suppose, instead of assuming all prices are given, and ordered as in
(\ref{priceorder}), it is now our decision to come up with the $m-1$
``discount'' prices, $p_2\ge \cdots \ge p_m$. The original
price $p_1$ is still given, and $p_1\ge p_2$. (The case when $p_1$
is also a decision variable is discussed at the end of this
section.)

Here, $m$ is also assumed to be given. That is, we limit the
number
of price takedowns that can take place over the horizon.
Furthermore, we assume the time horizon is divided into $m$
equal segments, with one price for each segment.
Without loss of generality, assume $T=m$, hence each segment is
of unit length.
Note that the equal-length segments do not contradict the earlier
models, where the time lengths $y_\ell$ corresponding to
accommodating
different prices are decision variables, and hence may vary in
$\ell$.
It can happen that in the derived pricing solution
we have $p_\ell= p_{\ell+1}$, for instance, then the price
$p_\ell$
will apply to two time segments instead of one.

Consider the model in
\S\ref{sec:independent}, i.e., batch Poisson order arrivals,
with the batch size independent of the price (hence,
we are not modeling phenomena such as quantity discount); however,
let the arrival rate be a decreasing function of the price,
$\g(p)$, which is also assumed to be differentiable.

Following the analysis that leads to the
problem formulation in (\ref{sobjmu}), and noticing that
$p_{1i}$ becomes $p_i$ here, since now there is only a single
price
in each time segment, we have  the
following optimization problem:
\begin{eqnarray}
\label{maxobj}
\max &&p_{1}[W - G(\mu_1)] + p_{2} [G(\mu_1)-G(\mu_2)]
 +\cdots + p_{m}[G(\mu_{m-1})-G(\mu_m)]\nonumber\\
&=&p_1W_1 - (p_1-p_2)G(\mu_1) -\cdots -(p_{m-1}-p_m) G(\mu_{m-1})-
p_m G(\mu_m),
\end{eqnarray}
where $(p_2,...,p_m)$ are decision variables, satisfying the
ordering in (\ref{priceorder}); and
$$\mu_i=\G_i=\g (p_1)+\cdots +\g (p_i), \qquad i=1,...,m.$$
Equivalently, we can solve the following minimization problem:
\begin{eqnarray}
\label{minobj}
\min_{p_2,\dots ,p_m} && (p_1-p_2)G(\mu_1) +\cdots +(p_{m-1}-p_m) G(\mu_{m-1})+
p_m G(\mu_m),\\
{\rm s.t.} && p_1\ge p_2\ge \cdots \ge p_m\ge 0.\nonumber
\end{eqnarray}

The above problem, although no longer separable,
can still be solved by a standard nonlinear (convex)
programming algorithm. Below, we present an approximation
algorithm that is easy to run and appears to work quite
well.

Apply a transformation of variable: let
the new decision variables be
\begin{eqnarray}
\label{p2r}
r_m := p_m; \qquad
r_i := p_i-p_{i+1},
\quad i=1,...,m-1.
\end{eqnarray}
The optimization problem in (\ref{minobj}) then becomes
as follows:
\begin{eqnarray}
\label{priceobj}
\min_r && \sum_{i=1}^m r_i G(\mu_i),\qquad {\rm s.t.}
\quad r_1+\cdots +r_m=p_1;
 \qquad r_i\ge 0, \quad i=1,...,m.
\end{eqnarray}
Write the derivatives
$$\g'_i:=\g'(p_i), \qquad \G'_i:=\g'_1+\cdots +\g'_i.$$
We can derive,
for any $i$ and $j$,
$$\frac{\partial\mu_i}{\partial r_j}=\G'_{i\wedge j}.$$
This follows from
noticing that  when $i>j$,
$p_i=r_i+\cdots+r_m$ does not involve $r_j$.

Therefore, the optimality equations are:
\begin{eqnarray}
\label{priceopteqn}
G(\mu_j)+\sum_{i=1}^m r_i G'(\mu_i)\G'_{i\wedge j}= \eta,
\qquad j=1,...,m .
\end{eqnarray}
Taking the difference between two consecutive equations above,
we have
\begin{eqnarray*}
G(\mu_j)-G(\mu_{j-1})+\g'_j\sum_{i=j}^m r_i G'(\mu_i)= 0,
\qquad j=2,...,m ;
\end{eqnarray*}
From the above, we can derive:
\begin{eqnarray}
\label{rj}
r_j= \frac{G(\mu_{j-1})-G(\mu_j)}{\g'_j G'(\mu_j)}
-\frac{1}{G'(\mu_j)}\sum_{i=j+1}^m r_i G'(\mu_i)
\qquad j=2,...,m .
\end{eqnarray}
Now, ignoring the second term on the right hand side above,
we have the following approximation:
\begin{eqnarray}
\label{rj1}
r_j= \frac{G(\mu_{j-1})-G(\mu_j)}{\g'_j G'(\mu_j)},
\qquad j=2,...,m .
\end{eqnarray}
Observe that
$$\g_j=\g(r_j+\cdots +r_m)=
\g(p_1-r_1-\cdots -r_{j-1})$$
only involves $(r_1,...,r_{j-1})$;
and hence, so does $\mu_j$.
That is, the right hand side of (\ref{rj1}), for each $j$,
only involves $(r_1,...,r_{j-1})$.
Therefore, following the recursion, we can relate
$r_2,..., r_m$ all to $r_1$, and then derive $r_1$ through a
simple line search via the equation
\begin{eqnarray}
\label{one}
r_1+r_2+\cdots +r_m=p_1.
\end{eqnarray}

In executing this recursion, we have noticed that when $p_1$ is
not large enough, $r_1$ might
become negative while we solve the equation
in (\ref{one}). This is because
the other $r_j$'s,
$j\neq 1$, are all positive, as evident from (\ref{rj1}).
One way to avoid this from happening is to first replace
$p_1$ by $Cp_1$, where $C$ is a large positive number.
Once all the $r_j$'s are derived and $r_1$ is positive,
divide each of these by $C$.


\medskip

In the following examples, we
consider three commonly used functions, the linear, exponential
and power
functions, that model the relation between the arrival rate and
the price:
\begin{eqnarray}
\label{pricefunc}
\g(p)=a-bp, \qquad \g(p)=ae^{-bp}, \qquad \g(p)=\frac{a}{p^b};
\end{eqnarray}
where $a$ and $b$ are positive parameters in all three cases.
The corresponding derivatives are:
$$\g'(p)=-b, \qquad \g'(p)=-b\g(p), \qquad \g'(p)=-
\frac{b}{p}\g(p).$$


\begin{exm}
\label{exm:pricing} {\rm In this example we choose $p_1=1$, and
consider the
linear, exponential and power functions in (\ref{pricefunc})
with different parameters $(a,b)$, along with different $W$
values.
The choice of the parameters $(a,b)$ is such that all three
functions coincide at $p=1$;
the linear and exponential functions also coincide at $p=0$;
while the power
function coincides with the linear function at a point close to
the
origin.

We list both the optimal solutions
and the approximations following (\ref{rj1}).
In each case, we consider both $m=8$ and $m=3$.
The results are summarized in Table \ref{tblpricingopt},
where listed under ``obj.\ val.''
are the original, {maximal} objective values following
(\ref{maxobj})
(in particular, including the $p_1W$ term).
}
\end{exm}

\begin{table}
\centering
\begin{tabular}{||c|c|cccccccc|c||}\hline 
$(W;a,b)$& $i=$ & 1& 2& 3& 4& 5& 6& 7& 8 & {obj.\ val.}\\
\hline
(40; 15, 14) & $p_i$  & 1  & 0.69 & 0.69&
0.69 & 0.69& 0.69 &0.68 & 0.62 & 25.64 \\
(40; 40, 37.33) & $p_i$ & 1  & 0.63 & 0.61&
& &  & &  & 22.94 \\
 \hline
(40; 15, 2) & $p_i$ & 1  & 0.60 & 0.60&
0.60&  0.60&0.60 & 0.59 &0.56  & 20.77 \\
(40; 40, 2) & $p_i$  & 1  & 0.55 & 0.54&
& &  & &  & 19.95 \\
\hline
(40; 2, 1.5) & $p_i$ & 1  & 0.57 & 0.57&
0.57&  0.57&0.57 & 0.54 &0.37  & 20.46 \\
(40; 5.33, 1.5) & $p_i$  & 1  & 0.52 & 0.44&
& &  & &  & 20.10
\\ \hline
\end{tabular}
\caption{\label{tblpricingopt}
Optimal pricing under linear, exponential and power demand
functions.
}
\end{table}

\begin{exm}
\label{exm:pricing2}
{\rm
Next, we examine the optimal price reduction with respect to the
available inventory $W$. We take the exponential demand case
above,
fix $m=8$ and $(a,b)=(15,2)$, while changing $W$.
The results are displayed in Table \ref{tblw}.
}
\end{exm}

\medskip

\begin{table}
\centering
\begin{tabular}{||c|c|cccccccc|c||}\hline 
$W$& $i=$ & 1& 2& 3& 4& 5& 6& 7& 8 & {obj.\ val.}\\
\hline
50 & $p_i$  & 1  & 0.52 & 0.52&
0.52 & 0.52& 0.52 &0.52 & 0.52 & 21.28 \\
 \hline
30 & $p_i$  & 1  & 0.73 & 0.73&
0.73 & 0.73& 0.73 &0.71 & 0.61 & 19.35 \\
 \hline
25 & $p_i$ & 1  & 0.82 & 0.82& 0.82
&0.82 &0.81 & 0.78  &0.65  & 18.12 \\
 \hline
20 & $p_i$ & 1  & 0.93 & 0.93&
0.93&  0.93&0.92 & 0.86 &0.69  & 16.43 \\
\hline
15 & $p_i$  & 1  & 1 & 1& 1
&1 &1  &1 & 0.78 & 14.00 \\
\hline
10 & $p_i$  & 1  & 1 & 1& 1
&1 &1  &1 & 0.88 & 9.90 \\
\hline
\end{tabular}
\caption{\label{tblw}
Optimal pricing under various inventory levels.
}
\end{table}

\begin{exm}
\label{exm:pricing3} {\rm Finally, we examine the approximation
scheme in (\ref{rj1}), which solves the optimality equations
approximately. We take the above cases under $m=3$. We also
examine non-optimal alternatives that offer different levels of
discount at different periods or no discount at all. (For
instance, in the second case, $(1, 0.90, 0.90)$ indicates full
price in period 1, and 10\% discount in the remaining two
periods.) The results are summarized in Table \ref{tblnonopt},
where the last column is the percentage off the objective value
under optimal pricing.
}
\end{exm}

\medskip

\begin{table}
\centering
\begin{center}
\begin{tabular}{||c|c|ccc|c|c||}\hline 
$(W;a,b)$& $i=$ & 1& 2& 3& {obj.\ val.} & \% off opt\\
\hline
(40; 40, 37.33) & $p_i$ & 1  & 0.66 & 0.45 &  21.82 (approx) & 5\%
\\
 & $p_i$ & 1  & 1 & 1 &  8.00 (non-opt) & 65\% \\
 \hline
(40; 40, 2) & $p_i$  & 1  & 0.68 & 0.35& 19.10 (approx) & 4\% \\
 & $p_i$  & 1  & 0.90 & 0.90& 17.32 (non-opt) &13\% \\
\hline
(40; 5.33, 1.5) & $p_i$  & 1  & 0.68 & 0.36& 19.84 (approx) &1\%
\\
 & $p_i$  & 1  & 1 & 0.50& 18.21 (non-opt) &10\% \\
\hline
\end{tabular}
\caption{\label{tblnonopt}
Approximations and non-optimal pricing schemes.}
\end{center}
\end{table}

\medskip

From the above results, we observe that
\begin{itemize}
\item[(a)]
reducing the number of price markdowns from 7 ($m=8$) to 2 ($m=3$)
has a rather minor effect on the objective values;
\item[(b)]
with more inventory available for sale, price reduction becomes
more
substantial and starts
earlier, as expected;
\item[(c)]
the approximation
scheme in (\ref{rj1})
performs quite well in all three cases;
\item[(d)]
applying the optimal pricing
results in a substantial advantage over other ad-hoc schemes.
\end{itemize}

\medskip

The model and analysis discussed above
extend readily to the case where, in addition to the
discount prices, the original price
$p_1$ is also a decision variable.
The optimization problem now becomes
\begin{eqnarray}
\label{p1}
\max_{p_1,r} && p_1 W-\sum_{i=1}^m r_i G(\mu_i) \\
{\rm s.t.} && r_1+\cdots +r_m\ge p_1;
\qquad p_1\ge 0; \; r_i\ge 0, \; i=1,...,m. \nonumber
\end{eqnarray}
The Lagrangian is
$$ p_1 W-\sum_{i=1}^m r_i G(\mu_i)+\eta (r_1+\cdots +r_m-p_1),$$
with $\eta$ being the multiplier.
Therefore, the optimality equations in (\ref{rj})
still apply; and in addition, we have $\eta=W$
(from setting the partial derivative with respect to $p_1$
to zero).
Similarly, the recursion in (\ref{rj1}) also applies,
assuming $p_1$, as well as $r_1$, is given;
and $r_1$ is still obtained from the summation
constraint:
$r_1+\cdots +r_m = p_1$.
(That this constraint  must be binding
follows from the fact that the term,
$\sum_{i=1}^m r_i G(\mu_i)$, is increasing in
$(r_1,...,r_m)$, which can be directly verified.)
Finally, $p_1$ can be obtained from another line search
to enforce
$\eta=W$.


\section{Concluding Remarks}
\label{sec:conclude}

The models discussed here can be extended in a number of ways.
First, the demand function in the pricing model
as represented by the Poisson arrival
rate can depend on time (period), $i$, as well as on price $p_i$,
to take the form $\g(i, p_i)$. For instance, a discount that takes
place earlier might attract more (or, less) demand.
The results in \S\ref{sec:pricing} will continue to hold,
since the only change needed is from  $\mu_i=\g(p_i)$ to
$\mu_i=\g(i,p_i)$.

Second, the available inventory $W$ can be made a decision
variable too. This amounts to constructing a newsvendor problem 
on top of the switch-over or pricing models developed here.
It is readily verified that all the objective functions
(for maximization) involved are concave in $W$. Hence, with
the addition of replenishment and salvage costs, the resulting
newsvendor model can be routinely solved, once the optimal
switching points or optimal prices are obtained (for each given
$W$).

Third, in the switch-over policy, we can incorporate
a service measure such as the acceptance rate
for each price class $k$: recall, it will not be accepted until
the time
interval $(t_{k-1},t_k]$.
Consider the model in \S\ref{sec:independent}, for instance.
We know the total expected number of accepted units, over the time
interval $(t_{k-1},t_k]$ is $G(\mu_{k-1})-G(\mu_k)$, of which the
share
of class $k$ is proportion to $\lam_k/\Lam_k$.
Hence, the expected number of accepted units for class $k$ over
the
entire horizon $(0,T]$ is:
$$\sum_{\ell=k}^m [G(\mu_{k-1})-
G(\mu_k)]\frac{\lam_k}{\Lam_\ell}.$$ The above divided by the
expected number of class $k$ arrivals over the horizon, $\lam_k
T\ex (Q)$, is what we call average acceptance rate for class $k$,
denoted $\al_k$. We can hence derive $\al_k$ after the optimal
switch-over points are derived; or, include a minimal requirement
for $\al_k$ as a constraint.


\end{document}